%% file: final.tex
\documentclass[onecolumn,a4paper,draftclsnofoot]{IEEEtran}
\usepackage{graphicx}
\usepackage{latexsym}
\usepackage{amsmath}
\usepackage{amssymb}
\usepackage{bm}
\usepackage{psfrag}
\usepackage{amsmath,amssymb}
\usepackage{graphicx}

\newcommand{\ttv}{\textsf{v}}

\newcommand{\argmax}{\mathop{\arg\max}\limits}
\newcommand{\diff}[1]{\mathrm{d}#1}

\newtheorem{theorem}{Theorem}

\newtheorem{corollary}{Corollary}
\newtheorem{definition}{Definition}
\newtheorem{discussion}{Discussion}

\begin{document}

\title{Analysis of Error Floors of Non-Binary LDPC Codes over MBIOS Channel}

\author{\IEEEauthorblockN{
Takayuki Nozaki, Kenta Kasai, and Kohichi Sakaniwa} \\
\IEEEauthorblockA{
Dept.\ of Communications and Integrated
Systems\\Tokyo Institute of Technology\\
152-8550 Tokyo, JAPAN\\
\{nozaki, kenta,  sakaniwa\}@comm.ss.titech.ac.jp
}
}
\maketitle

\begin{abstract}
In this paper, we investigate the error floors of non-binary 
low-density parity-check (LDPC) codes 
transmitted over the memoryless binary-input output-symmetric (MBIOS) channels.
We provide a necessary and sufficient condition for successful decoding 
of zigzag cycle codes over the MBIOS channel by the belief propagation decoder.
We consider an expurgated ensemble of non-binary LDPC codes
by using the above necessary and sufficient condition,
and hence exhibit lower error floors.
Finally, we show lower bounds of the error floors 
for the expurgated LDPC code ensembles over 
the MBIOS channel.
\end{abstract}

\section{Introduction}
Gallager invented low-density parity-check (LDPC) codes \cite{gallager_LDPC}.
Due to the sparseness of the parity check matrices,
LDPC codes are efficiently decoded by the belief propagation (BP) decoder. 
Optimized LDPC codes can exhibit performance very close 
to the Shannon limit \cite{richardson01design}.

Davey and MacKay \cite{DaveyMacKayGFq} 
have found that non-binary LDPC codes can outperform binary ones.
In this paper, we consider the non-binary LDPC codes defined over 
the Galois field $\mathbb{F}_{q}$ with $q=2^m$.

A non-binary LDPC code $C$ over $\mathbb{F}_q$ is defined 
by the null space of a sparse $M\times N$ parity-check matrix $H=(h_{i,j})$ 
over $\mathbb{F}_q$:
\begin{align*}
 C&=\Bigl\{\boldsymbol{x}\in \mathbb{F}_q^N
\mid H\boldsymbol{x}=\mathbf{0}\in\mathbb{F}_q^M\Bigr\}.
\end{align*}
The Tanner graph for a non-binary LDPC code is represented by
a bipartite graph with variable nodes, check nodes and labeled edges.
The $v$-th variable node and the $c$-th check node are connected 
with an edge labeled $h_{c,v}\in\mathbb{F}_q\setminus\{0\}$ if $h_{c,v}\ne 0$.
The LDPC codes defined by Tanner graphs with the variable nodes of degree $j$
and the check nodes of degree $k$ 
are called $(j, k)$-regular LDPC codes. 
It is empirically known that $(2,k)$-regular non-binary LDPC codes
exhibit good decoding performance among other LDPC codes for $q\ge 64$
\cite{Hu03regularand}. 
However, this is not the case for $q<64$. 
In this paper, we consider the irregular non-binary LDPC codes
which contain variable nodes of degree two for the generality of
code ensemble.

A zigzag cycle is a cycle such that
the degrees of al the variable nodes in the cycle are two.
In order to reduce the error floors of codes under maximum likelihood decoding,
Poulliat et al.\ proposed \textit{cycle cancellation} \cite{4641893}.
The cycle cancellation
is a method to design the edge labels in zigzag cycles so that
the corresponding submatrices are nonsingular.
We see that from the simulation results \cite{4641893}
the resulting codes have lower error floors under BP decoding.
However, it is found in our analyses that 
some zigzag cycles, even if their submatrices are nonsingular,
degrade decoding performance.
In this paper, we analyze a condition for successful decoding 
of zigzag cycles under BP decoding 
over the memoryless binary-input output-symmetric (MBIOS) channel.
Based on this condition, we propose a design method of selecting labels
so as to eliminate small zigzag cycles which degrade decoding performance.

In \cite{Noz_isit2010}, 
we analyze the decoding erasure rate in the error floors 
of non-binary LDPC codes 
over the binary erasure channel (BEC) under BP decoding.
In this paper, we analyze the error floors of non-binary LDPC codes
over the MBIOS channel.
First, we expurgate non-binary LDPC code ensembles.
Next, we show lower bounds for the symbol error rates in 
the error floors of the
expurgated LDPC code ensembles over the MBIOS channel.
More precisely, those lower bounds are derived from the decoding errors caused 
by the zigzag cycles.
Furthermore, simulation results show that the lower bounds on 
symbol error rates are tight for the expurgated ensembles constructed by
our proposed method over the MBIOS channels.

This paper is organized as follows. 
In Section \ref{sec:pre}, 
we briefly review the BP decoder
for non-binary LDPC codes.
In Section \ref{sec:zigzag_ana}, we clarify 
a necessary and sufficient condition for successful decoding 
of zigzag cycle codes over the MBIOS channel by the BP decoder
and propose a design method to lower the error floor.
In Section \ref{sec:ef_ana}, we give lower bounds 
for the symbol error rates in the error floors for
code ensembles constructed by our proposed method
and show that the proposed method gives
better performance than the cycle cancellation \cite{4641893}
and the method which uses both the cycle cancellation and 
the stopping set mitigation \cite{4641893}.

\section{Preliminaries \label{sec:pre}}
In this section, we recall the BP decoder 
for the non-binary LDPC codes \cite{DaveyMacKayGFq}.
We introduce some notations used throughout this paper.

\subsection{Channel Model}
Let $\alpha$ be a primitive element of $\mathbb{F}_{2^m}$.
Once a primitive element of $\alpha$ is fixed,
each element in $\mathbb{F}_{x^m}$
is given an $m$-bit representation \cite[p.~110]{macwilliams77}.
We denote the $m$-bit representation for $\gamma \in \mathbb{F}_{2^m}$,
by $(\gamma_1, \gamma_2, \dots, \gamma_m)$.
Let $\boldsymbol{x} = (x_1,x_2,\dots,x_N)$ 
denote the codeword over $\mathbb{F}_{2^m}$.
Since each symbol of $\mathbb{F}_{2^m}$ is given an $m$-bit representation,
a codeword is represented as a binary codeword of length $N m$,
$\boldsymbol{x} = (x_{1,1},x_{1,2},\dots,x_{N,m})$.

We denote the received word by $\boldsymbol{y} = (y_{1,1},y_{1,2},\dots,y_{N,m})$.
A channel is called {\it memoryless} binary-input channel if
\begin{align*}
 p(\boldsymbol{ y}\mid \boldsymbol{x}) 
  = 
 \prod_{i=1}^{N}\prod_{j=1}^{m} p(y_{i,j}\mid x_{i,j}).
\end{align*}
It is convenient to transform the binary input alphabet $\{0,1\}$
into $\{+1, -1\}$ by a binary phase shift keying (BPSK).
With some abuse of notation, we make no distinction between
$\{0,1\}$ and $\{+1,-1\}$.
A memoryless binary-input channel is called {\it output-symmetric} if
\begin{align*}
 p(y\mid x) =  p(-y\mid -x). 
\end{align*}
We assume the transmission over the MBIOS channel.
The MBIOS channels are characterized by its $L$-density $\textsf{a}$ \cite{mct}.
Examples of the MBIOS channels include the BEC,
the binary symmetric channel (BSC) 
and the additive white Gaussian noise (AWGN) channel.

\subsection{BP Decoder for Non-Binary LDPC Codes}
BP decoding proceeds by sending messages along the edges in the
Tanner graph.
The messages arising in the BP decoder for LDPC codes over $\mathbb{F}_{2^m}$
are vectors of length $2^m$.
Let $\Psi_{v, c}^{(\ell)}$ be the message from the $v$-th variable node
to the $c$-th check node at the $\ell$-th iteration.
Let $\Phi_{c, v}^{(\ell)}$ be the message from the $c$-th check node
to the $v$-th variable node at the $\ell$-th iteration.

\subsubsection{Initialization}
Set $\ell = 0$.
Let $N$ and $M$ be the number of variable nodes and check nodes 
in a Tanner graph,
respectively.
For $v =1,2,\dots,N$, 
let $C_{v} =(C_{v}(0), C_{v}(1),\dots,C_{v}(\alpha^{2^m-2}))$ 
denote the initial message of the $v$-th variable node.
For $\gamma \in \mathbb{F}_{2^m}$, 
the element of initial message $C_{v}(\gamma)$ 
is given from the channel output as follows: 
\begin{equation*}
 C_{v}(\gamma) 
  = 
 \prod_{i=1}^{m} {\rm Pr}\bigl(Y_{v,i} = y_{v,i} \mid X_{v,i} = \gamma_{i} \bigr).
\end{equation*}
Let $\mathcal{N}_{ \textsf{c} }(c)$ be the set of the positions of 
the variable nodes connecting to the $c$-th check node.
Set for all $c = 1,2,\dots,M$ and $v \in \mathcal{N}_{\textsf{c}}(c)$,
\begin{align*}
\Phi_{c,v}^{(0)} = \bigl(2^{-m},2^{-m},\dots,2^{-m}\bigr).
\end{align*}
\subsubsection{Iteration}
\paragraph{Variable node action}
Let $\mathcal{N}_{\ttv}(v)$ 
be the set of the positions of the check nodes 
connected to the $v$-th variable node.
The message $\Psi_{v,c}^{(\ell)}$ is given by the component-wise multiplication
of the initial message $C_{v}$
and the incoming messages $\Phi_{c',v}^{(\ell)}$ 
from check nodes whose positions $c'$ are in $\mathcal{N}_{\ttv}(v)$, 
i.e.,
for $x\in\mathbb{F}_{2^m}$
\begin{equation*}
 \Psi_{v,c}^{(\ell)}(x) 
  =
 \frac{1}{\xi} C_{v}(x) 
 \prod_{c'\in\mathcal{N}_{\ttv}(v)\setminus\{c\}} \Phi_{c',v}^{(\ell)}(x),
\end{equation*}
where $\xi$ is the normalization factor such that
$ 1
  =
 \sum_{x\in\mathbb{F}_{2^m}}  \Psi_{v,c}^{(\ell)}(x)$.

\paragraph{Check node action}
The convolution of two vectors $\Psi_1$ and $\Psi_2$ is given by
\begin{equation*}
 [\Psi_1 \oplus \Psi_2](x)
  =
 \sum_{y,z\in \mathbb{F}_{2^m} : x=y+z}
 \Psi_1(y)\Psi_2(z),
\end{equation*}
where $ \sum_{y,z\in \mathbb{F}_{2^m} : x=y+z} \Psi_1(y)\Psi_2(z)$ 
is the sum of $\Psi_1(y)\Psi_2(z)$ over all $y,z\in\mathbb{F}_{2^m}$
such that $x=y+z$.
To simplify the notation, we define 
$\bigoplus_{i\in\{1,2,\dots,k\}}\Psi_{i} 
:= \Psi_1 \oplus \Psi_2 \oplus \cdots \oplus \Psi_{k}$.

Let $h_{c,v}$ be the label of the edge adjacent to 
the $c$-th check node and the $v$-th variable node.
The message $\Phi_{c,v}^{(\ell+1)}$ is given as, for $x\in\mathbb{F}_{2^m}$
\begin{align*}
 &\check{\Psi}_{v,c}^{(\ell)}(x)
  =
 \Psi_{v,c}^{(\ell)}\Bigl(h_{c,v}^{-1} x\Bigr), \\
 &\check{\Phi}_{c,v}^{(\ell+1)} 
  =
 \bigoplus_{v'\in \mathcal{N}_{\textsf{c}}(c)\setminus \{v\}}
  \check{\Psi}_{v',c}^{(\ell)}, \\
 &\Phi_{c,v}^{(\ell+1)}(x)
  =
 \check{\Phi}_{c,v}^{(\ell+1)} (h_{c,v} x).
\end{align*}

\subsubsection{Decision}
Define 
\begin{align}
 \argmax_{x\in\mathbb{F}_{2^m}} \Psi 
  :=
 \bigl\{x\mid \forall y \in \mathbb{F}_{2^m}: \Psi(x) \ge \Psi(y) \bigr\},
\notag
\end{align}
and for $x\in\mathbb{F}_{2^m}$
\begin{equation}
 D_{v}^{(\ell)}(x)
  :=
 \frac{1}{\xi}C_{v}(x) \prod_{c\in\mathcal{N}_{\ttv}(v)} \Phi_{c,v}^{(\ell)}(x),
\notag
\end{equation}
where $\xi$ is the normalization factor such that 
$ 1 = \sum_{x\in\mathbb{F}_{2^m}}  D_{v}^{(\ell)}(x)$.
For $v = 1,2, \dots,N$,
let $\hat{x}_{v}^{(\ell)}$ be the decoding output of the $v$-th variable node.
Define
\begin{align}
\mathcal{D}_{v}^{(\ell)} 
  := 
\argmax_{x\in\mathbb{F}_{2^m}} D_{v}^{(\ell)}(x).
\notag
\end{align}
If $|\mathcal{D}_{v}^{(\ell)}| = 1$,
the decoding output $\hat{x}_{v}^{(\ell)}$ is the element of 
$\mathcal{D}_{v}^{(\ell)}$.
If $|\mathcal{D}_{v}^{(\ell)}| > 1$,
the decoder chooses 
$\hat{x}_{v}^{(\ell)} \in \mathcal{D}_{v}^{(\ell)}$
with probability $1/|\mathcal{D}_{v}^{(\ell)}|$.

\subsection{All-Zero Assumption and Defining Failure}
For the MBIOS channels, 
we assume that all-zero codeword is sent without loss of generality
to analyze the decoding error rate \cite{rathi_conditional}.

The $v$-th symbol is {\it eventually correct} \cite{R_ef}
if there exists $L_v$ such that for all $\ell > L_v$, $\hat{x}_{v}^{(\ell)}=0$.
The symbol error rate is defined by the fraction of the symbol 
which is not eventually correct.

\section{Condition for Successful Decoding over MBIOS Channel \label{sec:zigzag_ana}}

\begin{figure}[tb]
\begin{center}
 \includegraphics[height=40mm,clip]{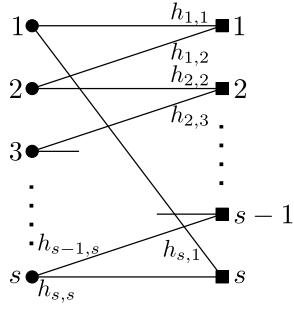} 
 \caption{A zigzag cycle code of symbol code length $s$. \label{fig:zigzag}}
\end{center}
\end{figure}

A zigzag cycle is a cycle such that 
the degrees of all the variable nodes in the cycles are two.
A zigzag cycle of weight $s$ consists of $s$ variable nodes of degree two.
The zigzag cycle code is defined by a Tanner graph which forms
a single zigzag cycle.
Figure \ref{fig:zigzag} shows a zigzag cycle code of symbol code length $s$.
In this section, we give a condition for successful decoding 
of the zigzag cycle codes over the MBIOS channels under BP decoding.

We consider the zigzag cycle code of symbol code length $s$ with labels
$h_{1,1}, h_{1,2}, \dots, h_{s,1}$ as shown in Fig.~\ref{fig:zigzag}.
We define $\gamma_{i} := h_{i,i}^{-1}h_{i,i+1}$ for $i=1,2,\dots,s$
where $h_{s,s+1} := h_{s,1}$.
Define $\beta := \prod_{i=1}^{s}\gamma_{i}$.
We refer to the parameter $\beta$ as \textit{cycle parameter}
\cite{Noz_isit2010}.
The following theorem shows
a necessary and sufficient condition for successful decoding 
of the zigzag cycle codes over the MBIOS channels 
by the BP decoder.
\begin{theorem} \label{the:1}
Let $\sigma$ be the order of $\beta$, i.e.,
let $\sigma$ be the smallest positive integer such that
$\beta^{\sigma}=1$.
We consider a zigzag cycle code of symbol code length $s$
defined over $\mathbb{F}_{2^m}$
with the cycle parameter $\beta$ over the MBIOS channel.
In the limit of large $\ell$,
all the symbols in the zigzag cycle code are eventually correct under BP decoding
if and only if
for all 
$x \in A_{\beta}:=\{\alpha^{j} \mid j = 0,1,\dots,\frac{2^m-1}{\sigma}-1\}$
\begin{equation}
 \prod_{k=1}^{s}(C_{k}(0))^{\sigma}
  >
 \prod_{t=0}^{\sigma-1}\prod_{k=1}^{s}
  C_{k}\left(\beta^{t}x\prod_{j=1}^{k-1}\gamma_{j}\right). 
 \notag
\end{equation}
Moreover,
in the limit of large $\ell$,
no symbols in the zigzag cycle code are eventually correct under BP decoding
if and only if
there exists
$x \in A_{\beta}:=\{\alpha^{j} \mid j = 0,1,\dots,\frac{2^m-1}{\sigma}-1\}$
such that
\begin{equation}
 \prod_{k=1}^{s}(C_{k}(0))^{\sigma}
  \le
 \prod_{t=0}^{\sigma-1}\prod_{k=1}^{s}
  C_{k}\left(\beta^{t}x\prod_{j=1}^{k-1}\gamma_{j}\right). 
 \notag
\end{equation}
 
\end{theorem}
\begin{IEEEproof}
 First, we write the messages $D_{v}^{(\ell)}$ by the initial messages $C_{v}$
 for the zigzag cycle code of symbol code length $s$ 
 with the cycle parameter $\beta$.
 Let $\tilde{\Psi}_{v,c}^{(\ell)}$ be the {\it unnormalized} message 
 from the $v$-th variable node to the $c$-th check node 
 at the $\ell$-th iteration.
 For all $x\in \mathbb{F}_{2^m}$ and $i=1,2,\dots,s$, 
 the unnormalized message for the zigzag cycle code of symbol code length $s$ 
 is written as follows:
 \begin{align}
  &\tilde{\Psi}_{i,i-1}^{(0)}(x) 
   := C_{i}(x),
  \hspace{3mm}
  \tilde{\Psi}_{i,i-1}^{(\ell+1)}(x) 
   := C_{i}(x) \tilde{\Psi}_{i+1,i}^{(\ell)}\Bigl(\gamma_{i}^{-1}x\Bigr), \notag \\
  &\tilde{\Psi}_{i,i}^{(0)}(x) 
   := C_{i}(x),
  \hspace{5.7mm}
  \tilde{\Psi}_{i,i}^{(\ell+1)}(x) 
   := C_{i}(x) \tilde{\Psi}_{i-1,i-1}^{(\ell)}\bigl(\gamma_{i-1}x\bigr), \notag \\
  &\tilde{D}^{(\ell+1)}_{i}(x)
   :=
  C_{i}(x)\tilde{\Psi}_{i-1,i-1}^{(\ell)}\bigl(\gamma_{i-1}x\bigr)
 \tilde{\Psi}_{i+1,i}^{(\ell)}\Bigl(\gamma_{i}^{-1}x\Bigr), \notag
 \end{align}
where $\tilde{\Psi}_{0,0}^{(\ell)} = \tilde{\Psi}_{s,s}^{(\ell)}$,
$\tilde{\Psi}_{1,0}^{(\ell)} = \tilde{\Psi}_{s+1,s}^{(\ell)} = 
\tilde{\Psi}_{1,s}^{(\ell)}$,
$\tilde{\Psi}_{s+1,s+1}^{(\ell)} = \tilde{\Psi}_{1,1}^{(\ell)}$
and $\gamma_{0} = \gamma_{s}$.
Then, for zigzag cycle code, the messages $\Psi_{i,j}^{(\ell)}$ and
$D_{i}^{(\ell)}$ are written as follows:
\begin{align}
 \Psi_{i,j}^{(\ell)}(x) 
  =
 \frac{\tilde{\Psi}_{i,j}^{(\ell)}(x)}
 { \sum_{ x'\in\mathbb{F}_{2^m} }\tilde{\Psi}_{i,j}^{(\ell)}(x')}, \hspace{1mm}
 D_{i}^{(\ell)}(x) 
  =
 \frac{\tilde{D}_{i}^{(\ell)}(x)}
 { \sum_{ x'\in\mathbb{F}_{2^m} }\tilde{D}_{i}^{(\ell)}(x')}. \notag 
\end{align}
From the definition, we have
\begin{equation}
 \tilde{D}^{(\ell)}_{i}(x)
  =
 C_{i}(x)
 \prod_{k=1}^{\ell}
  \left\{
    C_{i-k}\left(x\prod_{j=1}^{k}\gamma_{i-j}\right)
    C_{i+k}\left(x\prod_{j=0}^{k-1}\gamma_{i+j}^{-1}\right)
  \right\}, \label{eq:D}
\end{equation}
where $C_{i+ns}(x) = C_{i}(x)$ and $\gamma_{i+ns} = \gamma_{i}$ 
for $n=0,\pm 1,\dots$.
Define $\chi_{i} = \prod_{j=1}^{i-1}\gamma_j$
and 
\begin{align}
B(x) := 
 \prod_{t=0}^{\sigma-1}\prod_{k=1}^{s}
  C_{k} \left(\beta^{t} x \prod_{j=k}^{s}\gamma_{j} \right). \notag
\end{align}
From Eq.~\eqref{eq:D}, 
we have for $i=1,2,\dots,n$
\begin{align}
 &\tilde{D}_{i}^{(\ell + s\sigma)}(x)
  =
   \bigl\{ B\bigl(\chi_{i}x\bigr) \bigr\}^{2}
  \tilde{D}_{i}^{(\ell)}(x).
 \notag
\end{align}
By using this equation, we have
\begin{align}
 D_{i}^{(\ell_{1}s\sigma+\ell_{2})}(0)
  =
 \frac{\tilde{D}_{i}^{(\ell_{2})}(0)}
  {\tilde{D}_{i}^{(\ell_{2})}(0)
   +\sum_{x\in A_{\beta}}\Bigl\{\frac{B(\chi_{i}x)}{B(0)}\Bigr\}^{2\ell_{1}}
     \sum_{t=0}^{\sigma-1} \tilde{D}_{i}^{(\ell_{2})} \bigl(x\beta^{t}\bigr) }.
 \notag
\end{align}
If $B(0)>B(x)$ for all 
$x\in A_{\beta} = \{\alpha^{j} \mid j = 0,1,\dots,\frac{2^m-1}{\sigma}-1\}$, 
for all $i=1,2,\dots,s$,
we have $\lim_{\ell\to\infty} D_{i}^{(\ell)}(0) = 1$,
i.e., the decoding is successful.
If there exists 
$x\in A_{\beta}$
such that $B(0)<B(x)$,
for all $i=1,2,\dots,s$
we have $\lim_{\ell\to\infty} D_{i}^{(\ell)}(0) = 0$,
i.e., no symbols are eventually correct.

Finally, we claim that no symbols are eventually correct,
if there exists $x\in A_{\beta}$ such that $B(0)=B(x)$.
Note that for all $t\ge1$ and $i\in\{1,2,\dots,s\}$,
\begin{align}
 \tilde{D}_{i}^{(s\sigma t)}\Bigl(\chi_{i}^{-1}x\Bigr) 
 =&
 \bigl\{ B(x) \bigr\}^{2t}C_{i}\Bigl(\chi_{i}^{-1}x\Bigr)
  \notag \\
 \tilde{D}_{i}^{(s\sigma t-1)}\Bigl(\chi_{i}^{-1}x\Bigr)
 =&
 \bigl\{ B(x) \bigr\}^{2t} \Bigr\{C_{i}\Bigl(\chi_{i}^{-1}x\Bigr) \Bigl\}^{-1}.
  \notag
\end{align}
Hence for $t\ge 1$ and $i\in\{1,2,\dots,s\}$
\begin{align}
 \tilde{D}_{i}^{(s\sigma t)}\Bigl(\chi_{i}^{-1}x\Bigr) 
 \tilde{D}_{i}^{(s\sigma t-1)}\Bigl(\chi_{i}^{-1}x\Bigr)
 =&
  \bigl\{ B(x) \bigr\}^{4t} \notag \\
 =&
  \bigl\{ B(0) \bigr\}^{4t} \notag \\
 =&
 \tilde{D}_{i}^{(s\sigma t)}(0) 
 \tilde{D}_{i}^{(s\sigma t-1)}(0). \label{eq:equal_cond}
\end{align}
The $i$-th symbol is eventually correct if there exist $L$
such that $\tilde{D}_{i}^{(\ell)}(0) > \tilde{D}_{i}^{(\ell)}(x)$ 
for $\ell > L$ and $x\in\mathbb{F}_{2^m}$.
However, from Eq.~\eqref{eq:equal_cond},
for all $i\in\{1,2,\dots\}$,
if  
$\tilde{D}_{i}^{(s\sigma t-1)}(0) 
> \tilde{D}_{i}^{(s\sigma t-1)}(\chi_{i}^{-1}x)$,
then 
$\tilde{D}_{i}^{(s\sigma t)}(0) 
< \tilde{D}_{i}^{(s\sigma t)}(\chi_{i}^{-1}x)$.
Thus, no symbols are eventually correct.
This completes the proof.
\end{IEEEproof}

By Using Theorem \ref{the:1},
we have the following corollary.
\begin{corollary} \label{cor:1}
 Let $\sigma$ be the order of $\beta$.
 For a given channel output,
 if the zigzag cycle with cycle parameter $\beta$ 
 such that $\sigma\ne 2^m-1$
 is successfully decoded,
 then the zigzag cycle with cycle parameter $\beta$ 
 such that $\sigma = 2^m -1$
 is also successfully decoded.
\end{corollary}
\begin{IEEEproof}
We consider zigzag cycle of symbol code length $s$.
Since the channel output is given,
we are able to fix the initial message $C_i$ for $i=1,2,\dots,s$.
From Theorem \ref{the:1},
if the zigzag cycle with cycle parameter $\beta$ such that $\sigma \ne 2^m-1$
is successfully decoded,
then for all 
$x \in A_{\beta}=\{\alpha^{j} \mid j = 0,1,\dots,\frac{2^m-1}{\sigma}-1\}$
\begin{align*}
 \prod_{k=1}^{s}(C_{k}(0))^{\sigma}
  >
 \prod_{t=0}^{\sigma-1}\prod_{k=1}^{s}
  C_{k}\left(\beta^{t}x\prod_{j=1}^{k-1}\gamma_{j}\right).
\end{align*}
From the product of the above equation over all $x\in A_{\beta}$, we have
\begin{align*}
 &\prod_{x\in A_{\beta} }
 \prod_{k=1}^{s}(C_{k}(0))^{\sigma} 
>
 \prod_{x\in A_{\beta} }
 \prod_{t=0}^{\sigma-1}\prod_{k=1}^{s}
  C_{k}\left(\beta^{t}x\prod_{j=1}^{k-1}\gamma_{j}\right) \\
\iff&  
 \prod_{k=1}^{s}(C_{k}(0))^{2^m-1}
  >
 \prod_{x\in\mathbb{F}_{2^m}\setminus\{0\}}\prod_{k=1}^{s}
  C_{k}(x).
\end{align*}
From this condition, the zigzag cycle with cycle parameter $\beta$
such that $\sigma = 2^m-1$ is successfully decoded. 
\end{IEEEproof}

\begin{figure}[t]
\begin{center}
 \psfrag{beta}{$\beta$}
 \psfrag{a0}{$\alpha^0$} 
 \psfrag{a2}{$\alpha^2$} 
 \psfrag{a4}{$\alpha^4$}  
 \psfrag{a6}{$\alpha^6$} 
 \psfrag{a8}{$\alpha^8$} 
 \psfrag{a10}{$\alpha^{10}$} 
 \psfrag{a12}{$\alpha^{12}$}  
 \psfrag{a14}{$\alpha^{14}$} 
 \includegraphics[width=80mm,clip]{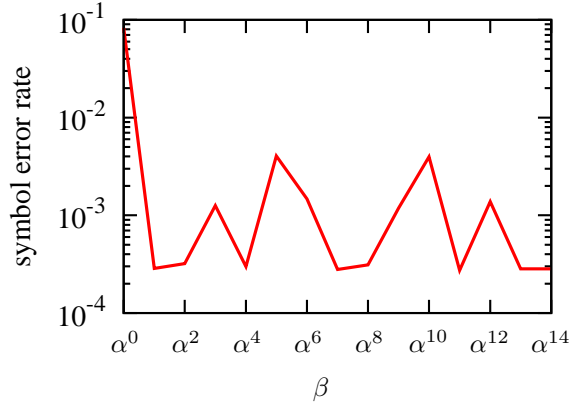} 
 \vspace{-1.5mm}
 \caption{The symbol error rate for the zigzag cycle code 
   define over $\mathbb{F}_{2^4}$ of symbol code length 3 
   over the AWGN channel with channel variance $\sigma^2=1$.
   The horizontal line corresponds to the cycle parameter.
 \label{fig:zigzag_comp}}
\end{center}
\end{figure}

\begin{discussion} \label{dis:1}
Corollary \ref{cor:1} shows that the zigzag cycles with cycle parameter
$\beta$ such that the order of $\beta$ is $2^m-1$ have
the best decoding performance.
We claim that the order of $\beta$ is $2^m-1$
if and only if $\beta \not\in \mathcal{H}_m$, where
\begin{equation}
 \mathcal{H}_{m}
  := \!
 \bigcup_{0<r<2^m-1:r\mid 2^m-1}\!
  \biggl\{\alpha^{i\frac{2^m-1}{r}}\mid i=0,\dots,r-1\biggr\}.
  \label{eq:H}
\end{equation}
Firstly, we show that the order of $\beta$ is $2^m-1$ 
if $\beta \not\in \mathcal{H}_m$.
For $r<2^m-1$, we define 
\begin{equation*}
  \mathcal{H}_{m}^{(r)} 
  := 
 \Bigl\{\alpha^{i\frac{2^m-1}{r}}\mid i=0,\dots,r-1\Bigr\}.
\end{equation*}
If $\beta\not\in\mathcal{H}_m^{(r)}$,
there exist integers $i\in\{0,1,\dots,r-1\}$ and $j\in\{1,\dots,(2^m-1)/r-1\}$ 
such that $\beta = \alpha^{i(2^m-1)/r+j}$.
Hence, we have
\begin{align*}
 \beta^{r}
  =
 \alpha^{\{i(2^m-1)/r+j\}r}
  =
 \alpha^{jr}.
\end{align*}
Since $jr<2^m-1$, we get $\beta = \alpha^{jr} \neq 1$.
Thus, we have the order of $\beta$ is not $r$ 
if $\beta \not\in \mathcal{H}_{m}^{(r)}$.
Since the order of $\beta$ is less than or equal to $2^m-1$ 
for $\beta\in\mathbb{F}_{2^m}\setminus\{0\}$,
the order of $\beta$ is $2^m-1$
if $\beta \not\in \mathcal{H}_m$.
Secondly, we show that $\beta \not\in \mathcal{H}_m$
if the order of $\beta$ is $2^m-1$.
Obviously, the order of $\beta\in \mathcal{H}_m^{(r)}$ is less than or equal 
to $r$.
Hence, the order of $\beta\in \mathcal{H}_m$ is less than $2^m-1$.
From the contraposition,
$\beta\not\in \mathcal{H}_m$
if the order of $\beta$ is $2^m-1$.
Therefore, we see that
the order of $\beta$ is $2^m-1$ if and only if $\beta \not\in \mathcal{H}_m$.

Thus, the zigzag cycles with the cycle parameter 
$\beta \not \in \mathcal{H}_{m}$ have the best decoding performance.
Note that $\{ \alpha^{i(2^m-1)/r} \mid i=0,\dots,r-1\}$ represents
a proper subgroup of $\mathbb{F}_{2^m}$.
Table \ref{tab:1} shows the elements in $\mathcal{H}_m$ for $m=2,3,4,5,6$.
Figure \ref{fig:zigzag_comp} shows the symbol error rate 
for the zigzag cycle code 
define over $\mathbb{F}_{2^4}$ of symbol code length 3 
over the AWGN channel with channel variance $\sigma^2=1$.
From Figure \ref{fig:zigzag_comp}, 
we see that the zigzag cycle codes with the cycle parameter 
$\beta\not\in\mathcal{H}_{4}$ have the best decoding performance.
\end{discussion}
\begin{table}[t]
  \begin{center}
   \caption{The elements in $\mathcal{H}_m$ for $m=2,3,4,5,6$.
   \label{tab:1}}
   \begin{tabular}{|c|l|}\hline
 Field & The elements of ${\mathcal{H}_m}$ \\ \hline 
 & \\ [-8pt]
 $\mathbb{F}_{2^2}$ & 1 \\ 
 & \\ [-8pt] \hline
 & \\ [-8pt]
 $\mathbb{F}_{2^3}$ & 1 \\ 
 & \\ [-8pt] \hline
 & \\ [-8pt]
 $\mathbb{F}_{2^4}$ & 
  $1, \alpha^{3}, \alpha^{5}, \alpha^{6}, \alpha^{9}, \alpha^{10}, \alpha^{12}$\\
 & \\ [-8pt] \hline
 & \\ [-8pt]
 $\mathbb{F}_{2^5}$ & 1 \\ 
 & \\ [-8pt] \hline
 & \\ [-8pt]
 $\mathbb{F}_{2^6}$ &  
  \parbox{80mm}{
  $1, \alpha^{3},  \alpha^{6}, \alpha^{7}, 
  \alpha^{9},  \alpha^{12},  \alpha^{14},  \alpha^{15}, 
  \alpha^{18}, \alpha^{21}, \alpha^{24}, \alpha^{25}, 
  \alpha^{27}, \alpha^{28}, \alpha^{30}$, \\
  $\alpha^{33}, \alpha^{35}, \alpha^{36}, \alpha^{39}, \alpha^{42}, 
  \alpha^{45}, \alpha^{48}, \alpha^{49}, \alpha^{51}, 
  \alpha^{54},  \alpha^{56}, \alpha^{57}, \alpha^{60}$}
  \\ [-8pt] & \\ \hline
   \end{tabular}
  \end{center}
\end{table}

By using the log-likelihood ratio, 
Theorem \ref{the:1} is simplified for 
the zigzag cycle codes with the cycle parameter $\beta\not\in\mathcal{H}_m$
over the MBIOS channel.
\begin{corollary} \label{cor:2}
We consider the zigzag cycle codes of symbol code length $s$
with the cycle parameter $\beta\not\in\mathcal{H}_m$
over the MBIOS channel.
Let $Z_{v,i}(Y_{v,i})$ be the log-likelihood ratio 
corresponding to the $i$-th channel output in the $v$-th variable node,
i.e.,
\begin{equation*}
 Z_{v,i}(Y_{v,i}) 
  = 
 \log\frac{{\rm Pr}( Y_{v,i} \mid X_{v,i} = 1 )}
      {{\rm Pr}( Y_{v,i} \mid X_{v,i} = -1 )}.
 \notag
\end{equation*}
In the limit of large $\ell$, 
all the symbols in the zigzag cycle code are eventually correct if and only if
\begin{equation*}
 \sum_{v=1}^{s}\sum_{i=1}^{m}Z_{v,i}(Y_{v,i}) > 0.
\end{equation*}
Moreover,
in the limit of large $\ell$, 
no symbols in the zigzag cycle code are eventually correct if and only if
\begin{equation*}
 \sum_{v=1}^{s}\sum_{i=1}^{m}Z_{v,i}(Y_{v,i}) \le 0.
\end{equation*}
\end{corollary}
\begin{IEEEproof}
To simplify the notation, we define for $k\in\{1,2,\dots,s\}$ and 
$i\in\{1,2,\dots,m\}$
\begin{align*}
p_{k,i} :=& {\rm Pr}(Y_{k,i} \mid X_{k,i}=1),   \\
\bar{p}_{k,i} :=& {\rm Pr}(Y_{k,i} \mid X_{k,i}=-1). 
\end{align*}
Note that for $k\in\{1,2,\dots,m\}$
\begin{align*}
 C_{k}(0) &= \prod_{i=1}^{m}p_{k,i}, \\
 \prod_{t=0}^{2^m-2} C_{k}\left(\beta^{t}\prod_{s=1}^{k-1}\gamma_{s}\right)
 &=
 \prod_{x\in \mathbb{F}_{2^m}\setminus\{0\}} C_{k}(x) \\
 &=
 \prod_{i=1}^{m}p_{k,i}^{2^{m-1}-1} \bar{p}_{k,i}^{2^{m-1}}
.
\end{align*}
From Theorem \ref{the:1}, 
all the symbols in the zigzag cycle are eventually correct if and only if
\begin{align*}
 &\prod_{k=1}^{s}(C_{k}(0))^{2^m-1}
  >
 \prod_{t=0}^{2^m-2}\prod_{k=1}^{s}
  C_{k}\left(\beta^{t}x\prod_{j=1}^{k-1}\gamma_{j}\right) \\
\iff&
 \prod_{k=1}^{s} \prod_{i=1}^{m}p_{k,i}^{2^m-1}
  >
 \prod_{k=1}^{s} \prod_{i=1}^{m}p_{k,i}^{2^{m-1}-1} \bar{p}_{k,i}^{2^{m-1}} \\
\iff&
 \sum_{k=1}^{s}\sum_{i=1}^{m}\log \frac{p_{k,i}}{\bar{p}_{k,i}} 
  >
 0.
\end{align*}
Hence, we see that
all the symbols in the zigzag cycle code are eventually correct if and only if
$ \sum_{k=1}^{s}\sum_{i=1}^{m} Z_{k,i}(Y_{k,i})  > 0.$
Similarly,
no symbols in the zigzag cycle code are eventually correct if and only if
$ \sum_{k=1}^{s}\sum_{i=1}^{m} Z_{k,i}(Y_{k,i})  \le 0.$
\end{IEEEproof}

\begin{figure}[tb]
 \begin{center}
 \includegraphics[width=80mm,clip]{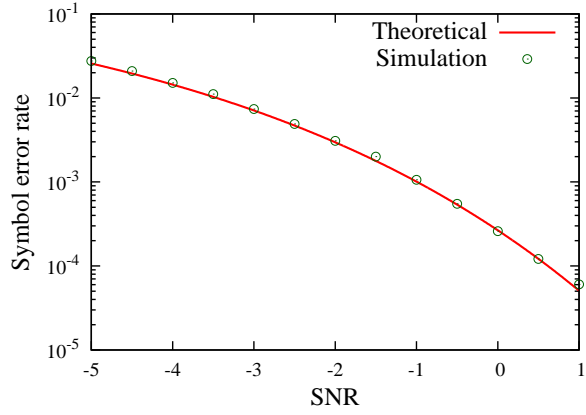} 
 \vspace{-1.5mm}
 \caption{Symbol error rate of zigzag cycle codes 
   defined over $\mathbb{F}_{2^4}$ of symbol code length 3
   with cycle parameter $\beta\not\in\mathcal{H}_4$.
   The continuous line shows the theoretical symbol error rate.
   The circles show the simulation result.
 \label{fig:zz_ana}}
 \end{center}
\end{figure}
Let ${\rm P}_{{\rm zz}}(s, m, \textsf{a})$
be the symbol error rate for
the zigzag cycle code defined over $\mathbb{F}_{2^m}$
of symbol code length $s$ with cycle parameter $\beta\not\in\mathcal{H}_m$
over the MBIOS channel characterized by its $L$-density $\textsf{a}$
under BP decoding.
Let $\textsf{a}_1,\textsf{a}_2,\dots,\textsf{a}_k$ denote 
independent and identically distributed random variables 
with density function $\textsf{a}$.
Define $Z^{(k)} := \sum_{i=1}^{k}\textsf{a}_i$.
From Corollary \ref{cor:2}, 
we have the symbol error rate of zigzag cycle code is
given by 
\begin{align}
 {\rm P}_{{\rm zz}}(s, m, \textsf{a})
  =
 {\rm Pr}(Z^{(sm)}\le 0).
\end{align}

Figure \ref{fig:zz_ana} shows the symbol error rate for
the zigzag cycle code defined over $\mathbb{F}_{2^4}$
of symbol code length $3$ with the cycle parameter $\beta\not\in\mathcal{H}_4$
over the AWGN channel.
The circles in Figure \ref{fig:zz_ana} show the simulation results.
The continuous line shows the theoretical symbol error rate.
For the AWGN channel with channel variance $\sigma$, 
the theoretical symbol error rate of the zigzag cycle codes
defined over $\mathbb{F}_{2^4}$ of symbol code length $s$
with cycle parameter $\beta\not\in\mathcal{H}_m$
is given by 
$ Q(\frac{\sqrt{sm}}{\sigma}),$
where 
$Q(y) = \frac{1}{\sqrt{2\pi}}\int^{\infty}_{y}\exp[-\frac{x^2}{2}] \diff{x}$.
From Figure \ref{fig:zz_ana}, we see that the the theoretical result
gives the symbol error rate of zigzag cycle code with the cycle parameter
$\beta\not\in\mathcal{H}_m$.





\section{Analysis of Error Floors \label{sec:ef_ana}}
In the previous section, we give a condition for the decoding error
to the zigzag cycle code.
By using this result, 
in this section, we give lower bounds of the symbol error rates
in the error floors of non-binary LDPC code ensembles over the MBIOS channel 
under BP decoding.

\subsection{Code Ensemble}
A stopping set $\mathcal{S}$ is a set of variable nodes
such that all the neighbors of $\mathcal{S}$ are connected to $\mathcal{S}$
at least twice.
Since the stopping sets depend only on the structure of a Tanner graph,
we are able to extend the definition of the stopping set for the non-binary LDPC codes.
Obviously, the zigzag cycles form stopping sets.

It is empirically known that $(2,k)$-regular non-binary LDPC codes
exhibit good decoding performance among other LDPC codes for $q\ge 64$
\cite{Hu03regularand}. 
However, this is not the case for $q<64$. In this paper, we consider the irregular non-binary LDPC codes
which contain variable nodes of degree two for the generality of
code ensemble.
Note that the (2,$k$)-regular non-binary LDPC code ensembles are included in
the irregular non-binary LDPC code ensembles which contain variable nodes of degree two.

From Discussion \ref{dis:1}, we see that the cycle parameter $\beta$
is an important parameter to improve the decoding error rate in the error floor.
The following definition gives expurgated ensembles 
parameterized by the cycle parameter $\beta$.
\begin{definition}  \label{def:3}
Let $\mathrm{LDPC}(N,m,\lambda,\rho)$ denote
LDPC code ensemble of symbol code length $N$ over $\mathbb{F}_{2^m}$
defined by Tanner graphs with a degree distribution pair $(\lambda, \rho)$ \cite{mct}
and labels of edges chosen elements from 
$\mathbb{F}_{2^m}\setminus \{0\}$ with equal probability.
Let $s_{\mathrm{g}}\in\mathbb{N}$ be an expurgation parameter.
The expurgated ensemble
$\mathrm{ELDPC}(N,m,\lambda,\rho,s_{\mathrm{g}})$ consists of the subset of
codes in $\mathrm{LDPC}(N,m,\lambda,\rho)$ which contain no stopping sets
of weight in $\{1,\dots,s_{\mathrm{g}}-1\}$.
Note that the expurgated ensemble $\mathrm{ELDPC}(N,m,\lambda,\rho,1)$ is
equivalent to $\mathrm{LDPC}(N,m,\lambda,\rho)$.
Let $s_{\mathrm{c}} \in \mathbb{N}$ 
be an expurgation parameter for labeling in the Tanner graph,
where $s_{\mathrm{g}} \le s_{\mathrm{c}}$.
Define expurgated ensemble
$\mathcal{E}(N, m, \lambda, \rho, s_{\mathrm{g}}, s_{\mathrm{c}},{\mathcal{H}})$
as the subset of codes in $\mathrm{ELDPC}(N,m,\lambda,\rho,s_{\mathrm{g}})$
which contain no zigzag cycles of weight in $\{s_{\mathrm{g}},\dots, s_{\mathrm{c}}-1\}$
with the cycle parameter $\beta\in\mathcal{H}$.
\end{definition}

Recall that $\alpha$ is a primitive element of $\mathbb{F}_{2^m}$.  
Define ${\mathcal{H}_m}$ as in Eq.~\eqref{eq:H}.
Note that the expurgated ensemble constructed by our proposed method 
and the cycle cancellation is represented as
$\mathcal{E}(N, m, \lambda, \rho, s_{\textsf{g}}, s_{\textsf{c}},\mathcal{H}_m)$
and 
$\mathcal{E}(N, m, \lambda, \rho, s_{\textsf{g}}, s_{\textsf{c}},\{1\})$,
respectively.

\subsection{Analysis of Error Floors}
In this section, we analyze the symbol error rate in the error floors
for the expurgated ensembles defined in Definition \ref{def:3}.
The following theorem gives a lower bound on the symbol error rate 
under BP decoding for the expurgated ensemble
$\mathcal{E}(N, m, \lambda, \rho, s_{\textsf{g}}, s_{\textsf{c}},\mathcal{H}_m)$.
\begin{theorem}\label{the:2}
Let ${\rm P}_{\rm s}(\mathcal{E},\textsf{a})$ 
be the symbol error rate of the expurgated ensemble
$\mathcal{E}(N, m, \lambda, \rho, s_{\textsf{g}}, s_{\textsf{c}}, \mathcal{H}_m)$
over the MBIOS channel characterized by its $L$-density $\textsf{a}$
under BP decoding.
Let $\textsf{a}_1,\textsf{a}_2,\dots,\textsf{a}_k$ denote 
independent and identically distributed random variables 
with density function $\textsf{a}$.
Define $Z^{(k)} := \sum_{i=1}^{k}\textsf{a}_i$ and $\mu := \lambda'(0)\rho'(1)$.
Let $\mathfrak{B}(\textsf{a})$ be the Battacharyya functional,
i.e., $\mathfrak{B}(\textsf{a}) := \int \textsf{a}(x) e^{-x/2}\diff{x}$.
The symbol error rate for sufficiently large $N$ and 
$\mathfrak{B}(\textsf{a}) < \mu^{-1/m}$
is bounded by
 \begin{align}
  {\rm P}_{{\rm s}}(\mathcal{E},\textsf{a})
 \ge
 \frac{1}{2N}\sum_{s=s_{\textsf{g}}}^\infty \mu^s {\rm Pr}\Bigl(Z^{(sm)}\le0\Bigr)
 +o\biggl(\frac{1}{N}\biggr).
 \label{eq:gen_ana}
 \end{align} 
\end{theorem}
\begin{IEEEproof}
Let $\tilde{P}(\mathcal{E},\textsf{a})$ be the symbol rate rate
caused by the zigzag cycles under BP decoding
over the MBIOS channel with characterized by its $L$-density $\textsf{a}$
for $\mathcal{E}(N, m, \lambda, \rho, s_{\textsf{g}},s_{\textsf{c}}, \mathcal{H})$.
Hence, we have
\begin{equation*}
{\rm P}_{{\rm s}}
   (\mathcal{E}, \textsf{a})
\ge
 \tilde{P}   (\mathcal{E}, \textsf{a}).
\end{equation*}

We will consider $\tilde{P}  (\mathcal{E}, \textsf{a})$.
Let $\tilde{P}_{1}(\mathcal{E}, \textsf{a},s)$ be the symbol error rate
caused by the zigzag cycles of weight $s$ under BP decoding
over the MBIOS channel with characterized by its $L$-density $\textsf{a}$
for $\mathcal{E}(N, m, \lambda, \rho, s_{\textsf{g}},s_{\textsf{c}}, 
{\mathcal{H}})$.
For the expurgated ensemble 
$\mathcal{E}(N, m, \lambda, \rho, s_{\textsf{g}},s_{\textsf{c}}, {\mathcal{H}})$,
the weights of zigzag cycles are at least $s_{\textsf{g}}$.
By \cite[C.~37]{mct}, if we fix a finite $W$ and let $N$ tend to infinity,
the zigzag cycles of weight at most $W$ become asymptotically non-overlapping
with high probability \cite[p.~155]{mct}.
Hence, for fixed $W$ and sufficiently large $N$ we have
\begin{align*}
 \tilde{P}(\mathcal{E},\textsf{a}) 
  \ge 
 \sum_{s=s_{\textsf{g}}}^{W} \tilde{P}_{1}(\mathcal{E},\textsf{a},s).
\end{align*}

From Corollary \ref{cor:2}, 
no symbols in zigzag cycle codes of weight $s$
with cycle parameter $\beta \not\in \mathcal{H}_{m}$ 
are eventually correct if $Z^{(sm)} \le 0$.
The symbol error rate for the zigzag cycle codes 
with $\beta \not\in \mathcal{H}_m$
is smaller than that for the zigzag cycle codes with $\beta \in \mathcal{H}_m$.
Hence, no symbols in the zigzag cycle of weight $s$
are eventually correct, with probability at least 
${\rm Pr}(Z^{(sm)} \le 0)$.
By \cite[C.~37]{mct} for fixed $W$,
the expected number of zigzag cycles of weight $s\le W$ 
is given by  $\mu^s/2s$,
for sufficiently large $N$.
Each zigzag cycle of weight $s$ causes a symbol error probability $s/N$.
Hence, for sufficiently large $N$, we have for 
$s\in\{s_{\textsf{g}},\dots,s_{\textsf{c}}-1\} $
\begin{align}
  \tilde{P}_{1}(\mathcal{E},\textsf{a},s)
 =
 \frac{1}{2N} \mu^s{\rm Pr}\Bigl(Z^{(sm)}\le 0\Bigr) 
 +o\biggl(\frac{1}{N}\biggr), 
\end{align}
and for $s\in\{s_{\textsf{c}},\dots, W\}$
\begin{align}
  \tilde{P}_{1}(\mathcal{E},\textsf{a},s)
 \ge
 \frac{1}{2N} \mu^s{\rm Pr}\Bigl(Z^{(sm)}\le 0\Bigr) 
 +o\biggl(\frac{1}{N}\biggr), 
\end{align}
Thus, we have
\begin{align}
 {\rm P}_{\textsf{s}}(\mathcal{E},\textsf{a})
 \ge
 \tilde{P}(\mathcal{E},\textsf{a})
 \ge&
 \frac{1}{2N}\sum_{s = s_{\textsf{g}}}^{W}  
 \mu^s{\rm Pr}\Bigl(Z^{(sm)}\le 0\Bigr)
 +o\biggl(\frac{1}{N}\biggr).
 \notag
\end{align}

Note that 
${\rm Pr}(Z^{(sm)}\le 0) 
 \le \bigl\{\mathfrak{B}(\textsf{a}) \bigr\}^{sm}$.
Hence, we have 
\begin{align}
 \frac{1}{2N}\sum_{s = s_{\textsf{g}}}^{W}  
 \mu^s{\rm Pr}\Bigl(Z^{(sm)}\le 0\Bigr)
  \le
 \frac{1}{2N}\sum_{s = s_{\textsf{g}}}^{W}  
 \Bigl(\mu \bigl\{\mathfrak{B}(\textsf{a})\bigr\}^{m} \Bigr)^s.
\notag
\end{align}
Thus, if $\mathfrak{B}(\textsf{a}) < \mu^{-\frac{1}{m}}$ and $W$ tends to infinity,
the left hand side of this inequality converges.
\end{IEEEproof}

\begin{corollary}
Define 
\begin{equation*}
 \epsilon^{*}_{m} := 
 \begin{cases}
  \frac{1}{2} & \text{for}~~ \mu \le 1, \\
  \frac{1-\sqrt{1-\mu^{-2/m}}}{2} & \text{for}~~ \mu > 1.
 \end{cases}
\end{equation*}
For the BSC with crossover probability $\epsilon$
and $\epsilon < \epsilon^{*}_{m}$,
the symbol error rate is lower bounded by
\begin{align}
 {\rm P}_{{\rm s}}(\mathcal{E},\textsf{a})
 \ge
 \frac{1}{2N}\sum_{s=s_{\textsf{c}}}^\infty \mu^s 
  \sum_{i\le ms/2}\binom{ms}{i}\epsilon^{ms-i}(1-\epsilon)^{i}
 +o\biggl(\frac{1}{N}\biggr).
  \label{eq:bsc_ana}
\end{align}
\end{corollary}

\begin{corollary}
Define 
\begin{equation}
 \sigma^{*}_{m} := 
 \begin{cases} 
  \infty & \text{for}~~ \mu \le 1, \\
  \sqrt{\frac{m}{2\ln \mu}} & \text{for}~~ \mu > 1.
 \end{cases} \notag
\end{equation}
For the AWGN channel 
with channel variance $\sigma^2$ and $\sigma < \sigma^{*}_{m}$,
the symbol error rate is lower bounded by
\begin{align}
 {\rm P}_{{\rm s}}(\mathcal{E},\textsf{a}) 
 \ge
\frac{1}{2N}\sum_{s=s_{\textsf{g}}}^\infty \mu^s 
  Q\Biggl(\frac{\sqrt{sm}}{\sigma}\Biggr) +o\Biggl(\frac{1}{N}\Biggr),
  \label{eq:awgnc_ana}
\end{align}
where 
$Q(y) = \frac{1}{\sqrt{2\pi}}\int^{\infty}_{y}\exp[-\frac{x^2}{2}] \diff{x}$.
\end{corollary}

\subsection{Simulation Results}
In this section, we compare the symbol error rate in the error floor 
for the expurgated ensemble constructed by our propose method
with (i) that constructed by the cycle cancellation \cite{4641893} 
in Section \ref{sec:AWGNC} and \ref{sec:BSC},
and (ii) that constructed by the combination of the cycle cancellation
and the stopping set mitigation \cite{4641893} in Section \ref{sec:ssm}.

\subsubsection{AWGN Channel Case \label{sec:AWGNC}}
\begin{figure}[tb]
\begin{center}
 \includegraphics[width=80mm,clip]{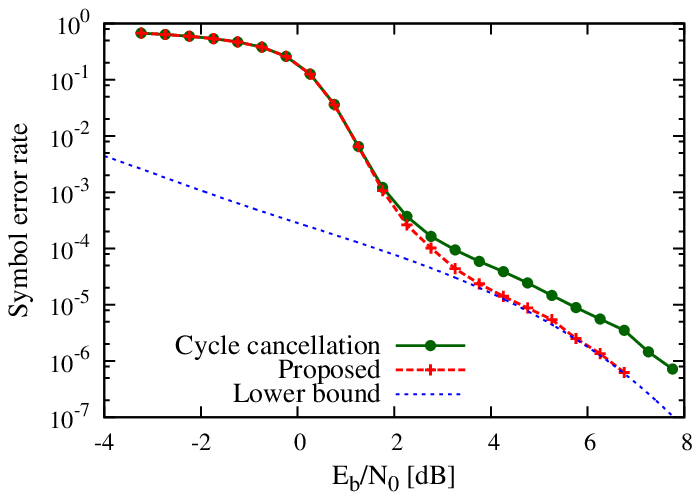} 
 \vspace{-1.5mm}
 \caption{Comparison of the symbol error rate for the expurgated ensemble
  $\mathcal{E}(315,4,x,x^2,1,8,\mathcal{H}_{4})$ (proposed) with 
  the expurgated ensemble 
  $\mathcal{E}(315,4,x,x^2,1,8,\{1\})$ (cycle cancellation).
  The lower bound is given by Eq.~\eqref{eq:awgnc_ana}.
 \label{fig:awgnc_reg}}
 \includegraphics[width=80mm,clip]{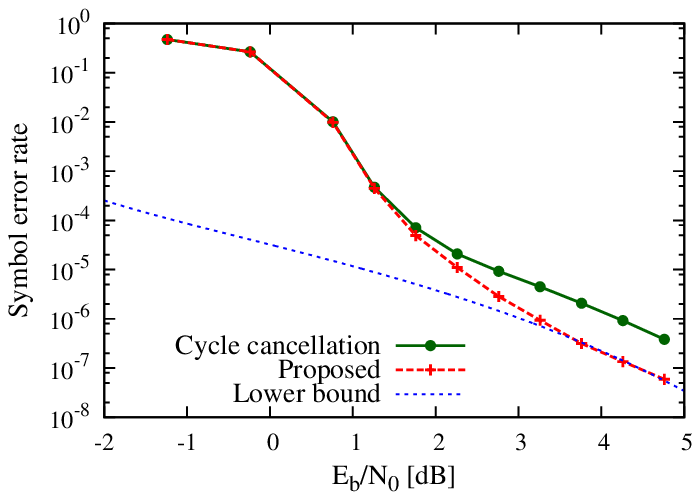} 
 \vspace{-1.5mm}
 \caption{Comparison of the symbol error rate for the expurgated ensemble
  $\mathcal{E}(1200,4,x,x^2,2,11,\mathcal{H}_{4})$ (proposed) with 
  the expurgated ensemble 
  $\mathcal{E}(1200,4,x,x^2,2,11,\{1\})$ (cycle cancellation).
  The lower bound is given by Eq.~\eqref{eq:awgnc_ana}.
 \label{fig:awgnc_ex}}
 \includegraphics[width=80mm,clip]{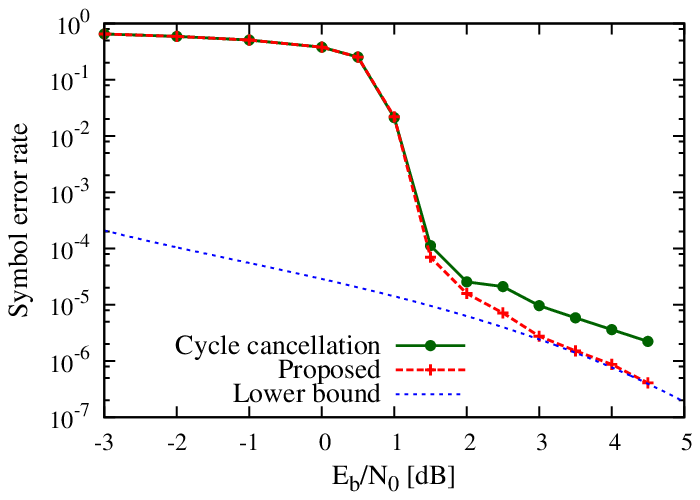} 
 \vspace{-1.5mm}
 \caption{Comparison of the symbol error rate for the expurgated ensemble
  $\mathcal{E}(1000,4,\lambda,\rho,1,8,\mathcal{H}_{4})$ (proposed) with 
  the expurgated ensemble 
  $\mathcal{E}(1000,4,\lambda,\rho,1,8,\{1\})$ (cycle cancellation),
  where $\lambda = 0.5x + 0.5x^2$ and $\rho = 0.5x^3 + 0.5x^5$.
  The lower bound is given by Eq.~\eqref{eq:awgnc_ana}.
 \label{fig:awgnc_ir}}
\end{center}
\end{figure}

First, we show the cases for regular non-binary LDPC code ensembles.
From Table \ref{tab:1}, we have 
$\mathcal{H}_{4} = 
\{1,\alpha^{3}, \alpha^{5},\alpha^{6}, \alpha^{9},\alpha^{10}, \alpha^{12}\}$.
Figure \ref{fig:awgnc_reg} and \ref{fig:awgnc_ex}
compare the symbol error rates for the expurgated ensembles
constructed by our proposed method 
$\mathcal{E}(315,4,x,x^2,1,8,\mathcal{H}_{4})$ and
$\mathcal{E}(1200,4,x,x^2,2,11,\mathcal{H}_{4})$ 
with the expurgated ensembles constructed by the cycle cancellation 
\cite{4641893} 
$\mathcal{E}(315,4,x,x^2,1,8,\{1\})$ and
$\mathcal{E}(1200,4,x,x^2,2,11,\{1\})$ 
respectively.
The lower bounds on symbol error rate are given by Eq.~\eqref{eq:awgnc_ana}.
Figure \ref{fig:awgnc_ex} is the case for $s_{\textsf{g}}>1$. 
We see that our proposed codes exhibit better decoding performance
than codes designed by the cycle cancellation.
We see that Theorem \ref{the:2} gives tight lower bounds for
the symbol error rates to the expurgated ensembles constructed by
our proposed method in the error floor.

Next, we show the case for an irregular non-binary LDPC code ensemble.
As an example, we employ the degree distribution pair
$\lambda = 0.5x + 0.5x^2$ and $\rho = 0.5x^3 + 0.5x^5$.
Figure \ref{fig:awgnc_ir} compares 
the symbol error rate for the expurgated ensembles
constructed by our proposed method 
$\mathcal{E}(1000,4,\lambda,\rho,1,8,\mathcal{H}_{4})$
with the expurgated ensembles constructed by the cycle cancellation.
$\mathcal{E}(1000,4,\lambda,\rho,1,8,\{1\})$.
The lower bounds on the symbol error rates
are given by Eq.~\eqref{eq:awgnc_ana}.
We see that our proposed codes exhibit better decoding performance
than codes designed by the cycle cancellation.
We see that Theorem \ref{the:2} gives a tight lower bounds for
the symbol error rates to the expurgated ensembles constructed by
our proposed method in the error floor.

\subsubsection{BSC Case \label{sec:BSC}}
\begin{figure}[tb]
\begin{center}
 \includegraphics[width=80mm,clip]{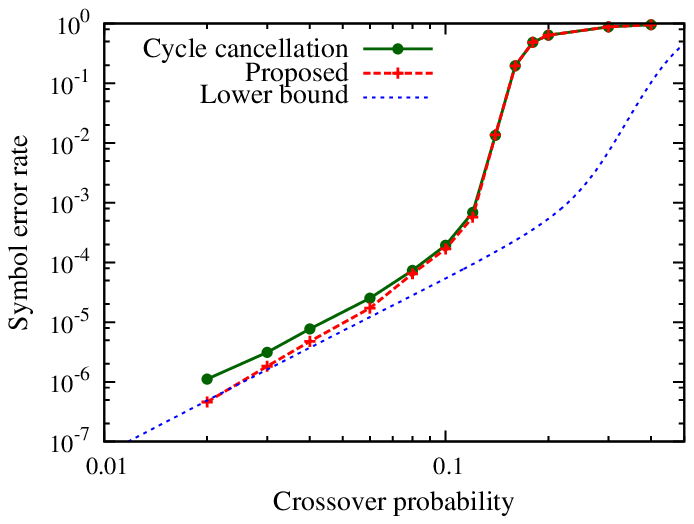} 
 \vspace{-1.5mm}
 \caption{Comparison of the symbol error rate for the expurgated ensemble
  $\mathcal{E}(315,6,x,x^2,1,8,\mathcal{H}_{6})$ (proposed) with 
  the expurgated ensemble 
  $\mathcal{E}(315,6,x,x^2,1,8,\{1\})$ (cycle cancellation).
  The lower bound is given by Eq.~\eqref{eq:bsc_ana}.
 \label{fig:bsc_reg}}
 \includegraphics[width=80mm,clip]{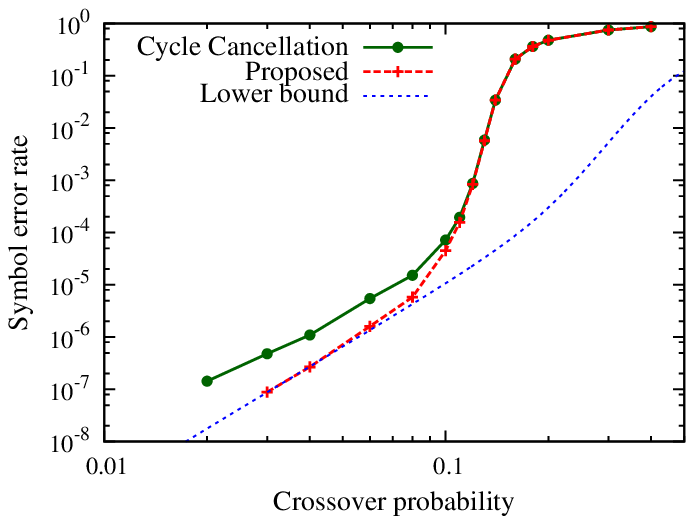} 
 \vspace{-1.5mm}
 \caption{Comparison of the symbol error rate for the expurgated ensemble
  $\mathcal{E}(1200,4,x,x^2,2,11,\mathcal{H}_{4})$ (proposed) with 
  the expurgated ensemble 
  $\mathcal{E}(1200,4,x,x^2,2,11,\{1\})$ (cycle cancellation).
  The lower bound is given by Eq.~\eqref{eq:bsc_ana}.
 \label{fig:bsc_ex}}
\end{center}
\end{figure}

Figure \ref{fig:bsc_reg} and \ref{fig:bsc_ex}
compare the symbol error rates for the expurgated ensembles
constructed by our proposed method 
$\mathcal{E}(315,6,x,x^2,1,8,\mathcal{H}_{6})$ and
$\mathcal{E}(1200,4,x,x^2,2,11,\mathcal{H}_{4})$ 
with the expurgated ensembles constructed by the cycle cancellation 
$\mathcal{E}(315,6,x,x^2,1,8,\{1\})$ and
$\mathcal{E}(1200,4,x,x^2,2,11,\{1\})$, 
respectively.
The lower bounds for the symbol error rates are given by Eq.~\eqref{eq:bsc_ana}.
Figure \ref{fig:bsc_ex} is
the case for $s_{\textsf{g}}>1$.
From Fig.\ \ref{fig:bsc_reg} and \ref{fig:bsc_ex}, 
we see that our proposed codes exhibit better decoding performance
than codes designed by the cycle cancellation.

We see that Theorem \ref{the:2} gives tight lower bounds for
the symbol error rates to the expurgated ensembles constructed by
our proposed method in the error floor.

\subsubsection{Comparison with Stopping Set Mitigation \label{sec:ssm}}
\begin{figure}[tb]
\begin{center}
 \includegraphics[width=80mm,clip]{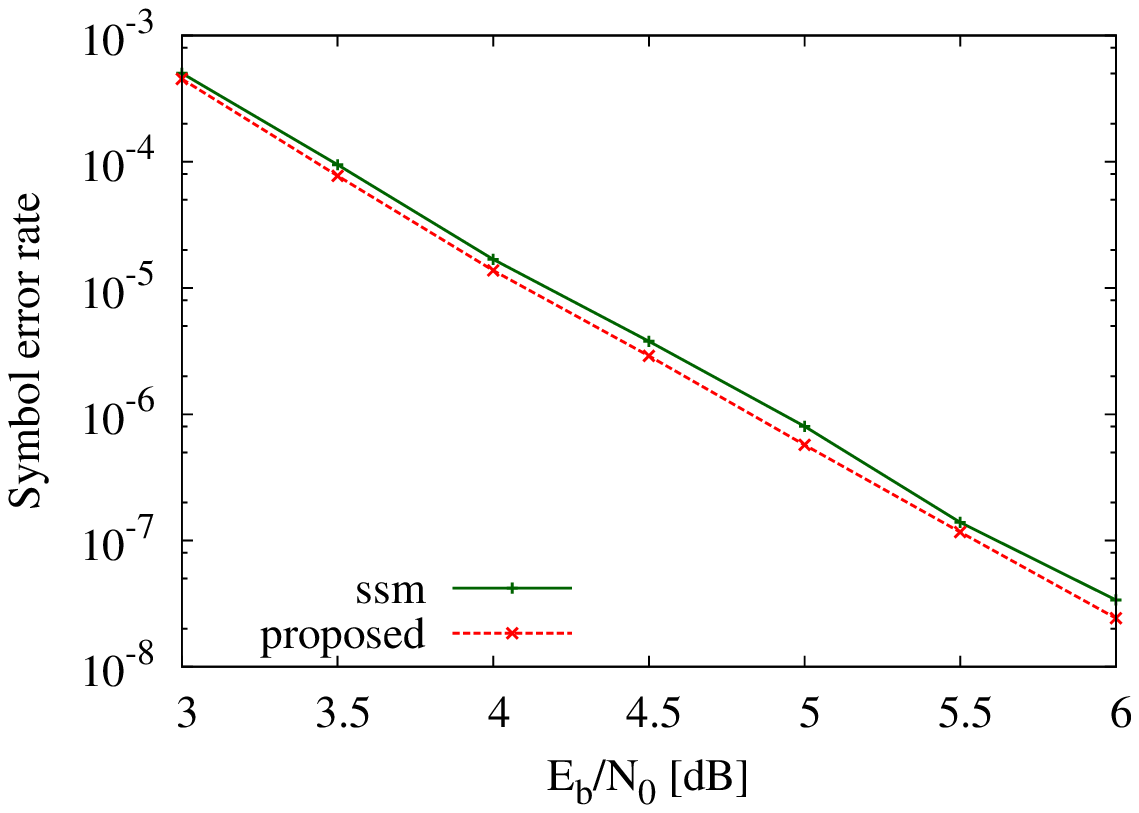} 
 \vspace{-1.5mm}
 \caption{Comparison of the symbol error rate for the codes
designed by the proposed method and
the codes designed by the method which uses both
the cycle cancellation and the stopping set mitigation.
The base code ensemble is $\mathrm{ELDPC}(60,4,x,x^3,3)$.
The curve (proposed) shows the symbol error rate for the codes designed by our proposed method.
The curve (ssm) shows the symbol error rate for the codes designed by
the method which uses both
the cycle cancellation and the stopping set mitigation.
 \label{fig:ssm60}}
\end{center}
  \end{figure}

In \cite{4641893}, Poulliat et al.\ also proposed 
the {\it stopping set mitigation}.
To lower the error floor further, Poulliat et al.\ proposed 
to use both the cycle cancellation and the stopping set mitigation.
We refer to the Hamming weight of the binary represented non-binary codeword
as binary weight.
the binary represented non-binary vectors.
The stopping set mitigation is a method to design the labels on the edges,
which are connecting to the nodes in the smallest stopping set,
so that the binary minimum distance in the stopping sets takes the maximum value.

Figure \ref{fig:ssm60} compares the symbol error rate for the codes
designed by the proposed method and
the codes designed by the method which uses both
the cycle cancellation and the stopping set mitigation \cite{4641893}.
In order to make the stopping set mitigation work effectively,
we employ as the base codes the codes whose Tanner graphs include many small stopping sets.
For example, this condition is met by the code ensemble $\mathrm{ELDPC}(60,4,x,x^3,3)$.
By applying our proposed method 
and the method which uses both the cycle cancellation and stopping set mitigation,
we get resulting codes which are the subsets of $\mathrm{ELDPC}(60,4,x,x^3,3)$.
We see Figure \ref{fig:ssm60} that the symbol error rate for our proposed method
is lower than that for the method using 
both the cycle cancellation and the stopping set mitigation.

\section{Conclusion}
In this paper,
we provide a necessary and sufficient condition for successful decoding 
of zigzag cycle codes over the MBIOS channel by the BP decoder.
Based on this condition, 
we propose a design method of selecting labels 
so as to eliminate small zigzag cycles which degrade decoding performance
for non-binary LDPC codes.
Finally, we show lower bounds of the error floors 
for the expurgated LDPC code ensembles over the MBIOS channel.

\section*{Acknowledgment}
This work was partially supported by Grant-in-Aid for JSPS Fellows.


\input{final.bbl}

\end{document}

%% file: final.bbl

%% file: final.bbl
\begin{thebibliography}{10}
\providecommand{\url}[1]{#1}
\csname url@samestyle\endcsname
\providecommand{\newblock}{\relax}
\providecommand{\bibinfo}[2]{#2}
\providecommand{\BIBentrySTDinterwordspacing}{\spaceskip=0pt\relax}
\providecommand{\BIBentryALTinterwordstretchfactor}{4}
\providecommand{\BIBentryALTinterwordspacing}{\spaceskip=\fontdimen2\font plus
\BIBentryALTinterwordstretchfactor\fontdimen3\font minus
  \fontdimen4\font\relax}
\providecommand{\BIBforeignlanguage}[2]{{%
\expandafter\ifx\csname l@#1\endcsname\relax
\typeout{** WARNING: IEEEtran.bst: No hyphenation pattern has been}%
\typeout{** loaded for the language `#1'. Using the pattern for}%
\typeout{** the default language instead.}%
\else
\language=\csname l@#1\endcsname
\fi
#2}}
\providecommand{\BIBdecl}{\relax}
\BIBdecl

\bibitem{gallager_LDPC}
R.~G. Gallager, \emph{Low {D}ensity {P}arity {C}heck {C}odes}.\hskip 1em plus
  0.5em minus 0.4em\relax in Research Monograph series, MIT Press, Cambridge,
  1963.

\bibitem{richardson01design}
T.~Richardson, M.~A. Shokrollahi, and R.~Urbanke, ``Design of
  capacity-approaching irregular low-density parity-check codes,'' \emph{{IEEE}
  Trans. Inf. Theory}, vol.~47, pp. 619--637, Feb. 2001.

\bibitem{DaveyMacKayGFq}
M.~Davey and D.~MacKay, ``Low-density parity check codes over {GF}($q$),''
  \emph{{IEEE} Commun. Lett.}, vol.~2, no.~6, pp. 165--167, Jun. 1998.

\bibitem{Hu03regularand}
X.-Y. Hu, E.~Eleftheriou, and D.~Arnold, ``Regular and irregular progressive
  edge-growth tanner graphs,'' \emph{{IEEE} Trans. Inf. Theory}, vol.~51,
  no.~1, pp. 386--398, Jan. 2005.

\bibitem{4641893}
C.~Poulliat, M.~Fossorier, and D.~Declercq, ``Design of regular
  (2,$d_c$)-{LDPC} codes over {GF}($q$) using their binary images,''
  \emph{{IEEE} Trans. Commun.}, vol.~56, no.~10, pp. 1626--1635, Oct. 2008.

\bibitem{Noz_isit2010}
T.~Nozaki, K.~Kasai, and K.~Sakaniwa, ``Error floors of non-binary {LDPC}
  codes,'' in \emph{Proc. 2010 {IEEE} Int. Symp. Inf. Theory(ISIT)}, Jun. 2010,
  pp. 729--733.

\bibitem{macwilliams77}
F.~J. MacWilliams and N.~J.~A. Sloane, \emph{The Theory of Error-Correcting
  Codes}.\hskip 1em plus 0.5em minus 0.4em\relax Amsterdam: Elsevier, 1977.

\bibitem{mct}
T.~Richardson and R.~Urbanke, \emph{Modern Coding Theory}.\hskip 1em plus 0.5em
  minus 0.4em\relax Cambridge University Press, Mar. 2008.

\bibitem{rathi_conditional}
V.~Rathi, ``Conditional entropy of non-binary {LDPC} codes over the {BEC},'' in
  \emph{Proc. 2008 {IEEE} Int. Symp. Inf. Theory(ISIT)}, Jul. 2008, pp.
  945--949.

\bibitem{R_ef}
T.~J. Richardson, ``Error floors of {LDPC} codes,'' in \emph{Proc. 41th Annual
  Allerton Conf. on Commun., Control and Computing}, Oct. 2003, pp. 1426--1435.

\end{thebibliography}
